\documentclass[xcolor=dvipsnames]{PoS}

\title{\vspace*{-1.5cm}
\begin{flushright}
\texttt{\footnotesize CERN-PH-TH-2015-040}
\end{flushright}
\vfill
Gauge-invariant signatures of spontaneous gauge symmetry breaking 
by the Hosotani mechanism}

\ShortTitle{Detecting gauge symmetry breaking}

\author{Oscar \AA kerlund \\
  Institut f\"ur Theoretische Physik, ETH Zurich, CH-8093 Z\"urich, Switzerland\\
  E-mail: \email{oscara@phys.ethz.ch}}

\author{\speaker{Philippe de Forcrand}\\
  Institut f\"ur Theoretische Physik, ETH Zurich, CH-8093 Z\"urich, Switzerland\\
  CERN, Physics Department, TH Unit, CH-1211 Geneva 23, Switzerland\\
  E-mail: \email{forcrand@phys.ethz.ch}}

\abstract{
The Hosotani mechanism claims to achieve gauge-symmetry breaking,
for instance $SU(3) \to SU(2)\times U(1)$. To verify this claim,
we propose to monitor the stability of a topological defect stable
under a gauge subgroup but not under the whole gauge group,
like a $U(1)$ flux state or monopole in the case above. We use gauge
invariant operators to probe the presence of the topological defect
to avoid any ambiguity introduced by gauge fixing.
Our method also applies to an ordinary gauge-Higgs system.
          }

\FullConference{The 32nd International Symposium on Lattice Field Theory,\\
		23-28 June, 2014\\
		Columbia University New York, NY}

\begin{document}

\section{Introduction}
Dimensional reduction tells us that QCD at high temperature can be 
effectively described as a $3d$ Yang-Mills theory, plus an adjoint Higgs
field generated by the static mode of $A_0$, i.e. by the Polyakov loop.
Dimensional reduction also occurs in the case of a compact {\em extra}
dimension: a $(4+1)d$ Yang-Mills theory is effectively described as
a $4d$ Yang-Mills, plus an adjoint Higgs field coming from the Polyakov
loop $P_5$ in the extra dimension. This led Hosotani \cite{Hosotani:1983}, in 1983, to
the scenario of ``gauge-Higgs unification'': by a judicious choice of
matter content and boundary conditions in the extra dimension, the minimum
of the effective potential for ${\rm Tr}P_5$ can be displaced from its
trivial value $A_5=0$. Then, the corresponding $4d$ adjoint Higgs field
acquires a non-trivial vacuum expectation value, which can [partially]
break the gauge symmetry of the Yang-Mills theory.
While this scenario seems to be disfavored phenomenologically, we are 
concerned here with a different aspect: how can one diagnose the claimed
breaking of {\em gauge} symmetry? Our proposal actually applies also 
to genuine (gauge + Higgs) systems.

\section{The Hosotani mechanism}
We consider a $(4+1)$-dimensional theory with gauge group $SU(3)$ and
a compact extra dimension. The effective potential $V_{\rm eff}({\rm Tr}P_5)$
can be obtained in perturbation theory, by considering the static modes
of the gauge and matter fields. As shown by Hosotani, adjoint fermions
with periodic boundary conditions in the extra dimension give a contribution
opposite to that of gluons, and can displace the minimum of $V_{\rm eff}$
from the trivial one (Fig.~1 left) to that of Fig.~1 middle or right.
The 3 cases have been dubbed ``deconfined'', ``split'' and ``reconfined''.
They exist also in a $(3+1)$-dimensional system, where they were found
in a serendipitous way~\cite{Cossu:2009sq}, in a project aimed at 
understanding center symmetry breaking in QCD(Adj) as proposed by Unsal~\cite{Unsal:2006pj}.

\begin{figure}
\centerline{
\includegraphics[width=0.28\textwidth]{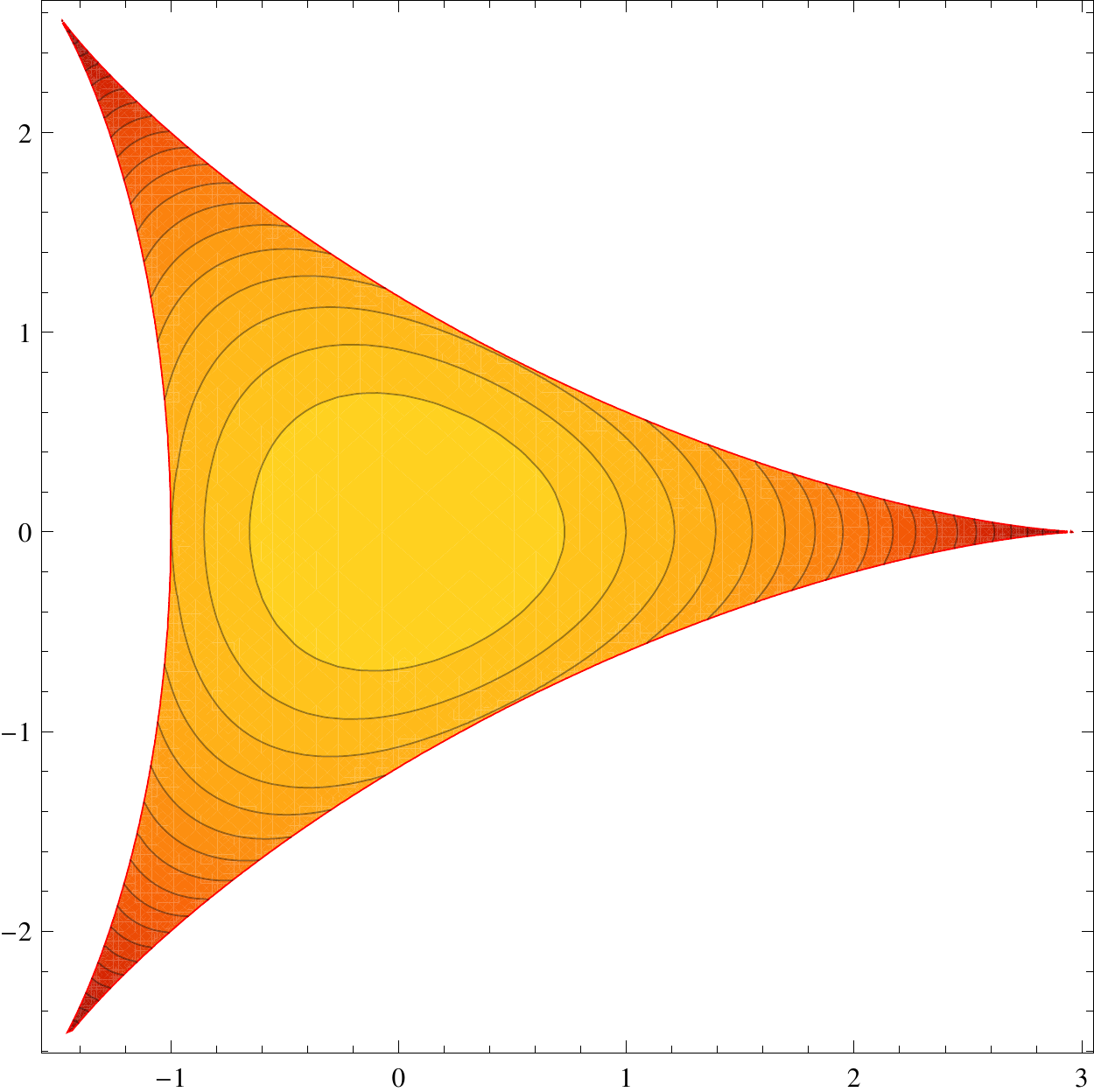}
\put(-116,114.5){$\bullet$}
\put(-4,59){$\bullet$}
\put(-116,3.5){$\bullet$}
\put(-55,20){deconfined}
\hspace*{2mm}
\includegraphics[width=0.28\textwidth]{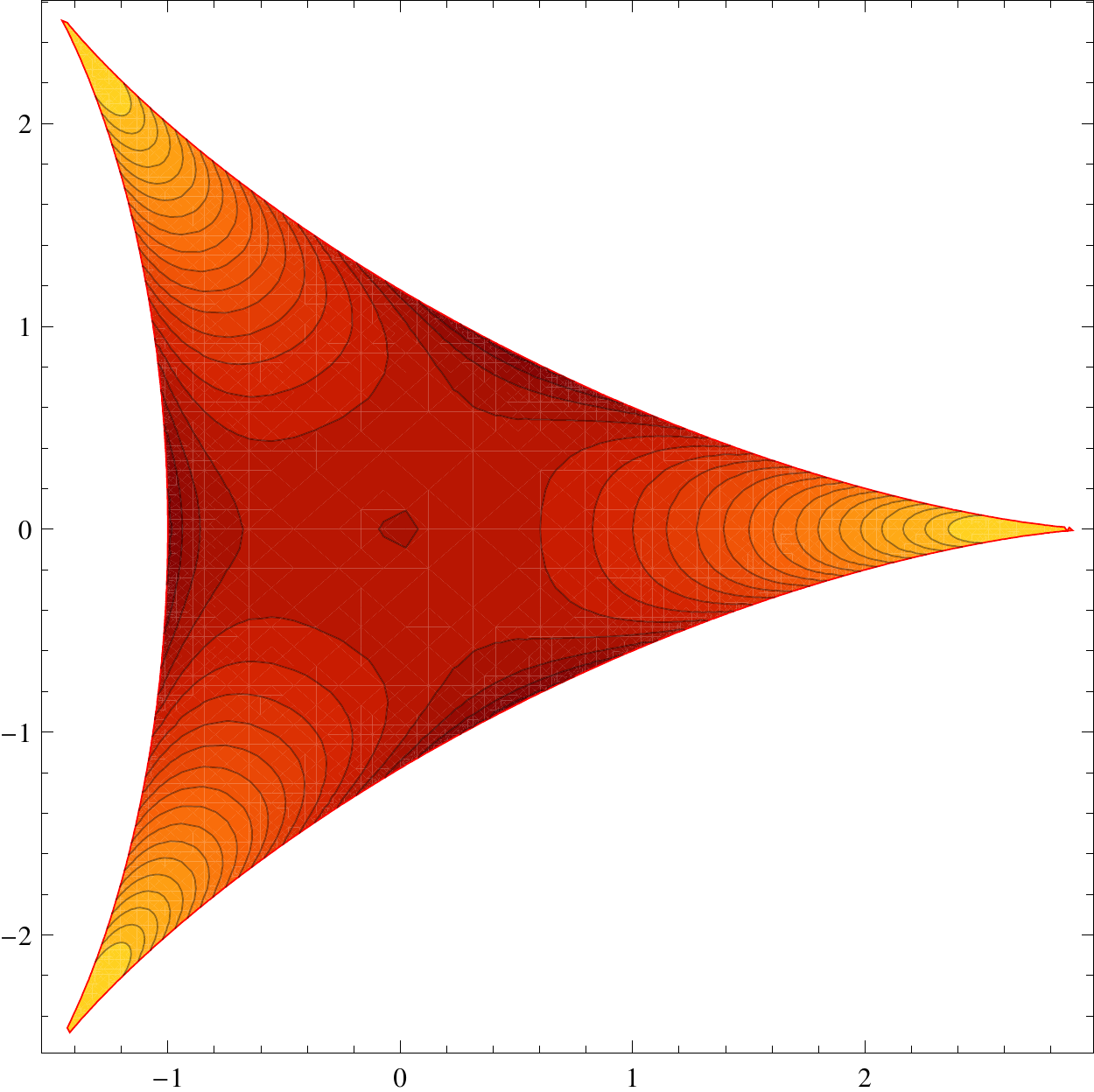}
\put(-65,40){$\bullet$}
\put(-65,78){$\bullet$}
\put(-105,59){$\bullet$}
\put(-55,20){split}
\hspace*{2mm}
\includegraphics[width=0.28\textwidth]{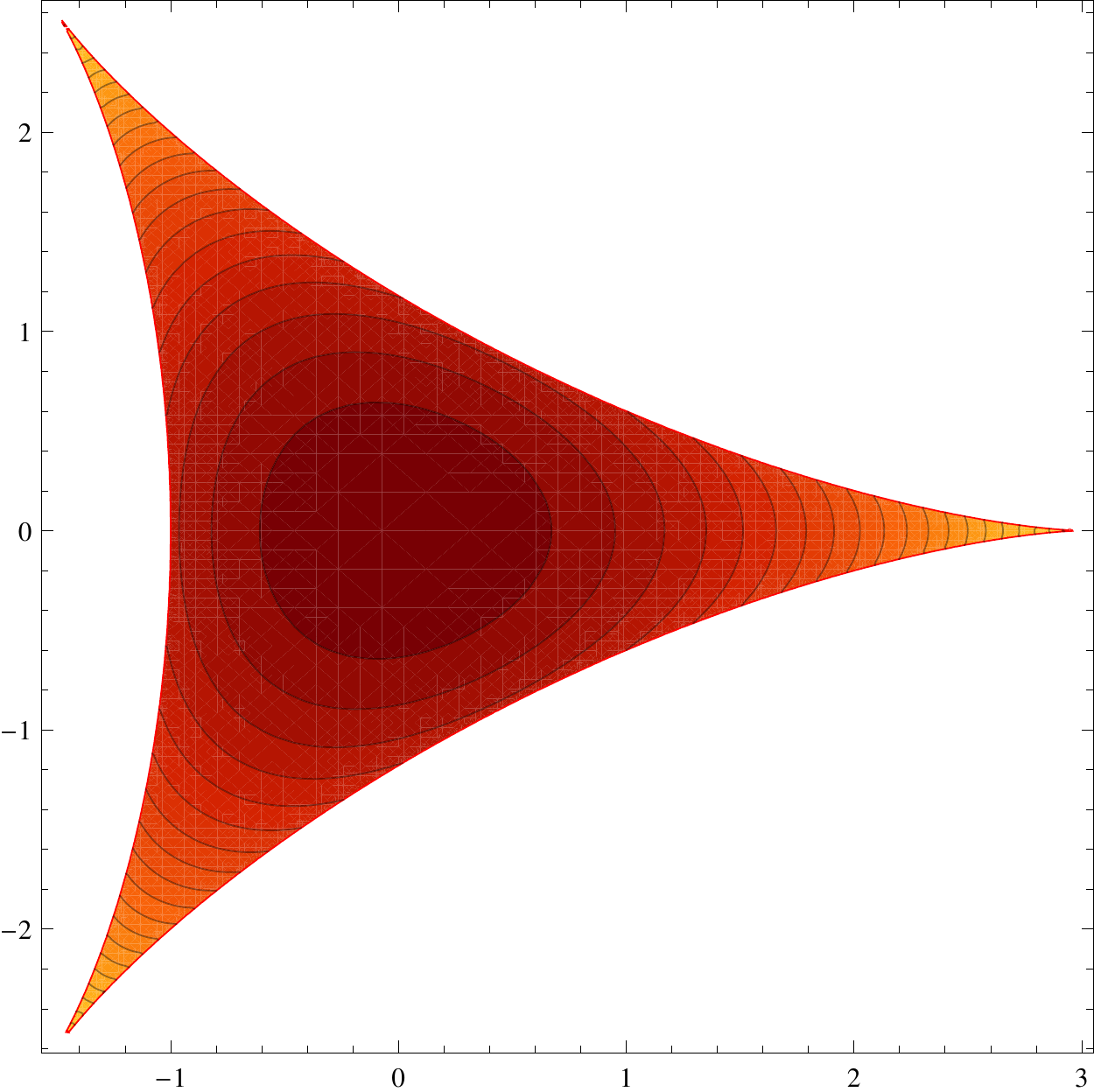}
\put(-81,59){$\bullet$}
\put(-55,20){reconfined}
}
\caption{Contour lines of the effective potential $V_{\rm eff}({\rm Tr} P_5)$ in the three phases, resulting from different matter contents and boundary conditions in the extra dimension.
Black dots mark the minima in each case.}
\end{figure}
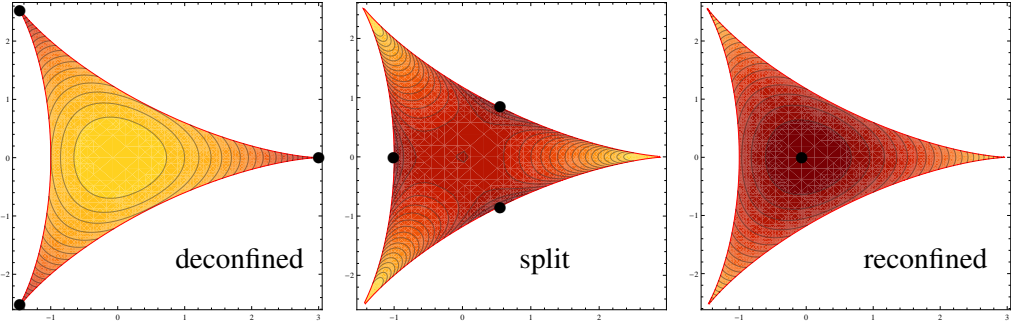

Let us consider the diagonalized form of the $SU(3)$ matrix $P_5$ in the 
3 cases. Up to eigenvalue permutations and global phases 
$\exp(\pm i\frac{2\pi}{3})$, one finds \\
$\left( \begin{array}{ccc} +1 & 0 & 0 \\ 0 & +1 & 0 \\  0 & 0 & +1 
\end{array} \right)$ (``deconfined''),
$\left( \begin{array}{cc|c} -1 & 0 & 0 \\ 0 & -1 & 0 \\ \hline 0 & 0 & +1 
\end{array} \right)$ (``split''), and
$\left( \begin{array}{cc|c} e^{+i\frac{2\pi}{3}} & 0 & 0  \\ 
0 & e^{-i\frac{2\pi}{3}} & 0 \\ \hline 0 & 0 & 1 \end{array} \right)$ 
(``reconfined'').
While the first matrix, the identity, is invariant under any $SU(3)$
gauge transformation $P_5 \leftarrow \Omega^\dagger P_5 \Omega$  $\forall
\Omega \in SU(3)$,
this is not true of the other two matrices.
In the ``split'' case, $P_5$ is left unchanged only if 
$\Omega \in SU(2)\times U(1)$.
In the ``reconfined'' case, $P_5$ is left unchanged only if 
$\Omega = \exp(i \theta_3 \lambda_3 + i \theta_8 \lambda_8)$, i.e.
$\Omega \in U(1)\times U(1)$.
Therefore, we are in a situation where the action maintains full $SU(3)$
gauge invariance, but the vacuum does not (in the split and reconfined cases):
this characterizes the breaking of the gauge symmetry.

The situation is no different from that of an ordinary Higgs field: one says 
that the gauge symmetry ``breaks'' when the Higgs field ``develops an
expectation value'', although the action remains gauge-invariant and 
the expectation value $\langle\phi \rangle$ remains exactly zero (in the
absence of gauge-fixing). Here too, $\langle A_5\rangle$ remains exactly
zero under the action of gauge transformations $\Omega$ which permute 
the eigenvalues of $P_5$.

Nevertheless, the long-distance physics of the 3 phases is dramatically
different: the $4d$ theory has 8 gluons in the first case, or 3 gluons
and 1 photon in the second case, or 2 photons in the third. 
The presence of photons gives rise to a Coulomb potential between 
corresponding probe charges. 
Yet, these different physics are very hard to recognize,  because the 
corresponding $SU(2)$ and $U(1)$ gauge subgroups are ``scrambled''
differently at each lattice site.

Let us focus on the ``split'' $SU(2)\times U(1)$ phase.
One might consider detecting the Coulomb potential by measuring $SU(3)$
Wilson loops. But this will not work: each link is a product of an $SU(2)$
and a $U(1)$ element, and the trace of such a loop will obey an area law
coming from the $SU(2)$ factors, regardless of the $U(1)$ factors.
Clearly, finding a signature of the above phenomena in the infrared 
properties of the effective $4d$ theory is not as simple as one might
initially think. One way out would perhaps be to fix the 
gauge~\footnote{No signal was found by Jim Hetrick, as presented in \cite{Hetrick:2014hqa}.},
with a gauge condition which minimizes the magnitude of the $SU(3)$ link
elements which do not belong to $SU(2)\times U(1)$ or $U(1)\times U(1)$.
The latter is the well-known Maximal Abelian gauge. 
Here, we want to study gauge-invariant observables, and follow a different
route.

\section{Gauge symmetry breaking seen in a gauge-invariant way}

Since the eigenvalues of $P_5$ are gauge-invariant, it is instructive to consider the quadratic fluctuations
$m_k^{-2} = \langle (\bar{A_5}^k - \langle \bar{A_5}^k\rangle)^2 \rangle$
of the static mode $\bar{A_5}$ defined via $P_5 = \exp(i g L_5 \bar{A_5}^k \lambda_k)$.
These fluctuations determine the Higgs mass squared. A possible $k$-dependence will be a clear
sign of gauge-symmetry breaking. These masses can be determined by perturbation theory
around the vacuum corresponding to each phase. They are shown in Fig.~2~\cite{Cossu:2013ora}.
As one would expect, fluctuations of $\bar{A_5}$ about the trivial vacuum in the deconfined phase
are isotropic in color space. The same is true in the reconfined phase, where $Z_3$ symmetry is 
enforced in $P_5$ locally. Remarkably, in the split phase two different masses are found, for
fluctuations of $\bar{A_5}$ in the $\lambda_3$ and $\lambda_8$ directions. 
In other words, $V_{\rm eff}(\bar{A_5})$ is elliptical, not spherical, in the split phase.
However, this interesting phenomenon is of no relevance to long-distance $4d$ physics,
because the Higgs mass is of order $1/L_5$. 

\begin{figure}
\centerline{
\includegraphics[width=0.55\textwidth]{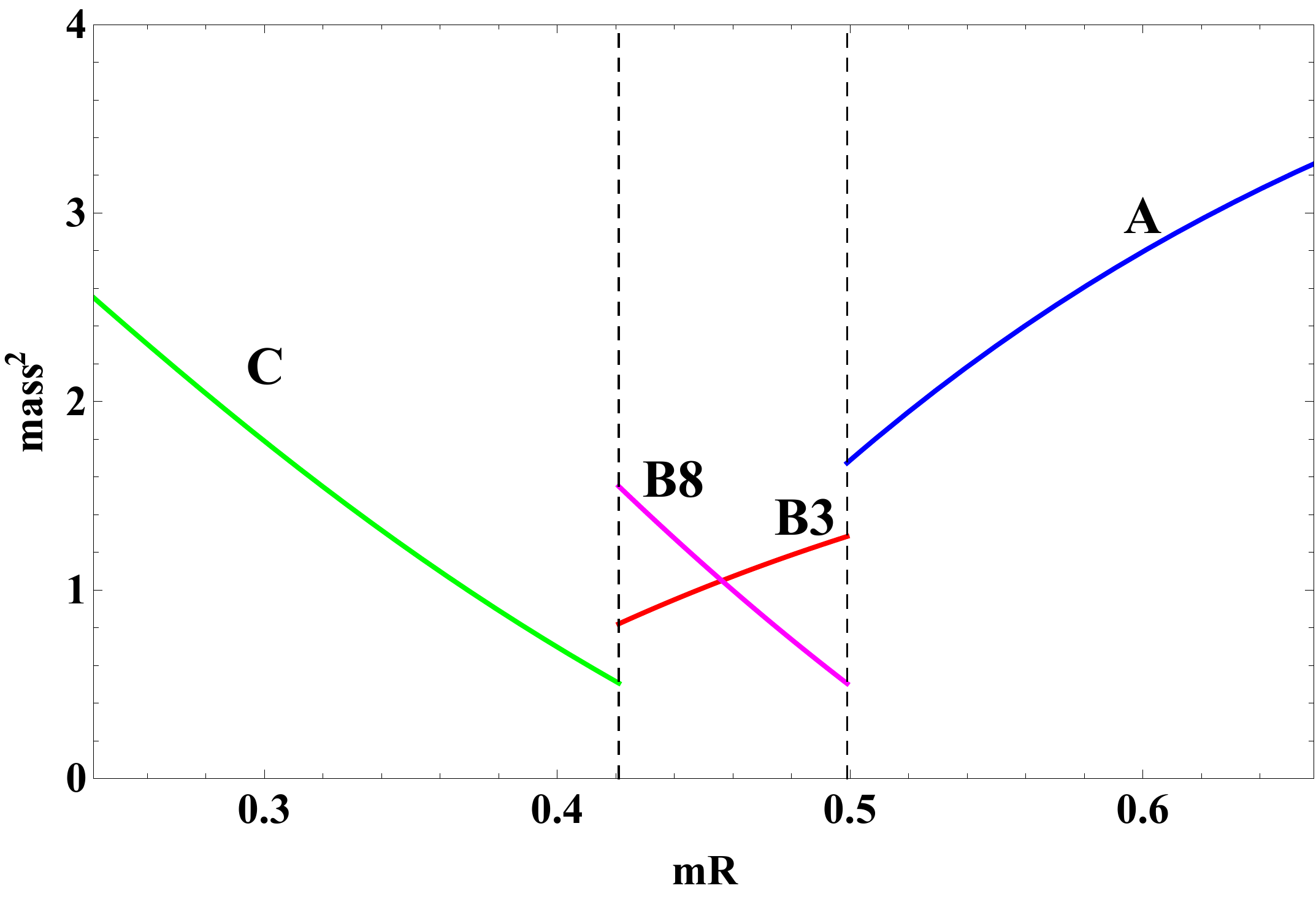}
\put(-65,85){\color{blue} deconfined}
\put(-115,85){\color{red} split}
\put(-185,85){\color{green} reconfined}
}
\caption{Mass squared of the Higgs field, as a function of the size of the extra dimension, for two adjoint fermions. In the split phase, the two Abelian components of the Higgs field have different masses. Perturbative calculation and figure from \cite{Cossu:2013ora}.}
\end{figure}

Our proposal is to monitor the stability of topological excitations supported by a gauge subgroup
in the case of gauge-symmetry breaking, but not by the whole gauge group.
The simplest example is provided by $\Pi_1(U(1)) = {\cal Z}$, while $\Pi_1(SU(3)) = \Pi_1(SU(2)) = \bf{1}$.
The corresponding topological excitation, an Abelian flux in some plane $xy$, is going to be
stable if the gauge symmetry is broken down to a subgroup $U(1)$, but will otherwise unwind
in a larger $SU(3)$ or $SU(2)$ gauge group. The same happens with an Abelian monopole.
The stability of these objects can be monitored via gauge-invariant observables like the plaquette.

Let us first consider an $xy$ Abelian flux in a $U(1)$ gauge theory on an $L_x \times L_y \times L_z \times L_t$ 
lattice.  To prepare such a state, start from a "cold" configuration $U_\mu(x)=\bf{1}~\forall x,\mu$.
Then, in each $xy$ plane, insert a $2\pi$ flux by arranging the links so that each $xy$ plaquette $P_{xy}$
is equal to $\exp(i\frac{2\pi}{L_x L_y})$. From this starting configuration, perform usual Monte Carlo
updates, and monitor the gauge-invariant flux action: 
$\Delta = \langle {\rm Tr}U_{P_{xz}} \rangle - \langle {\rm Tr}U_{P_{xy}} \rangle$,
where the difference is taken to isolate the effect of the $xy$ flux.
Classically, $\Delta = 1 - \cos\frac{2\pi}{L_x L_y} \approx \frac{2\pi^2}{L^2_x L^2_y}$.
The leading effect of fluctuations is to modify $P_{xy}$ and $P_{xz}$ in the same way,
so that $\Delta \sim \langle {\rm Tr}U_{P_{xz}} \rangle (1 - \cos\frac{2\pi}{L_x L_y} B)$
for $B$ units of flux. This simple prediction is completely consistent with the numerical
simulation of Fig.~3, where $B$ is incremented every 50 Monte Carlo sweeps.
Flux states are extremely stable, since for their decay one $xy$ plaquette in each plane
must go through angle $\pi$. This only happens at the right edge of the figure.

\begin{figure}
\centerline{
\includegraphics[width=0.55\textwidth]{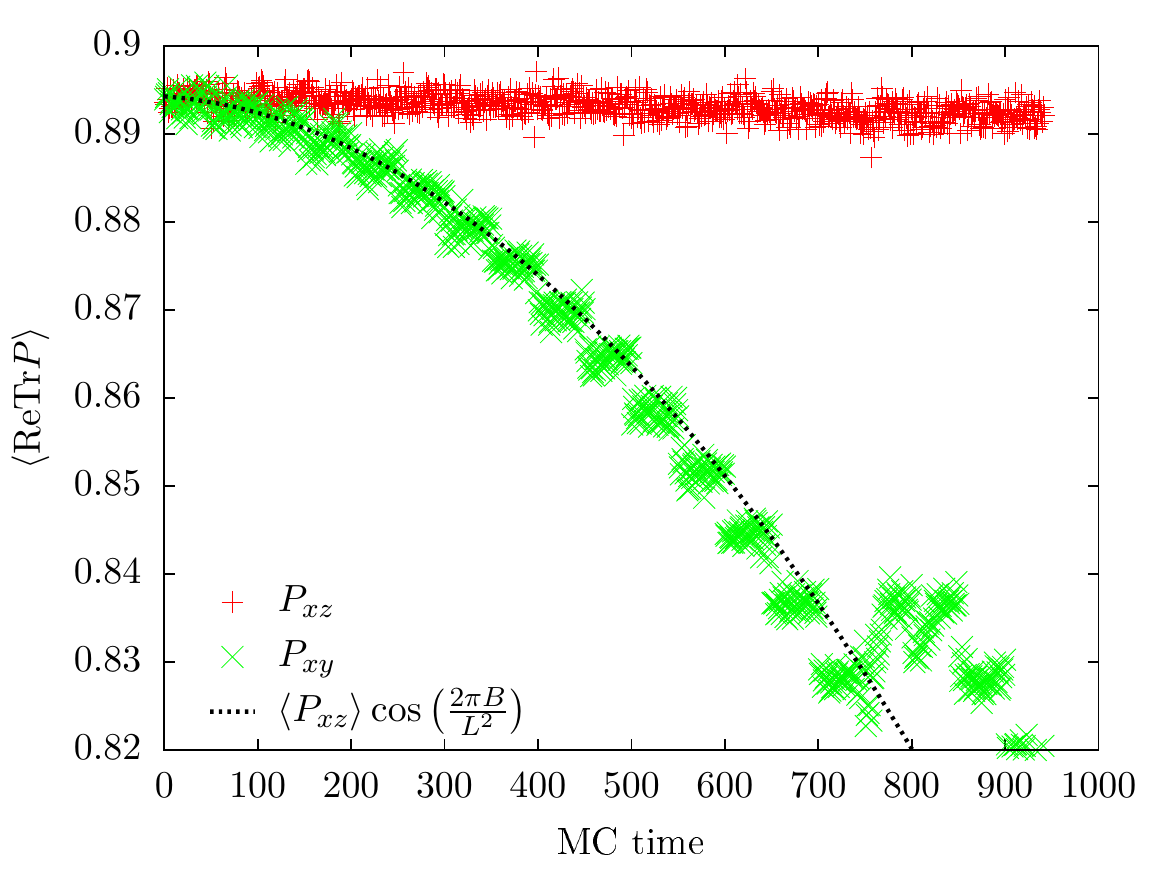}
}
\caption{Average $xy$ and $xz$ plaquettes in a $U(1)$ system, under the introduction of one unit of $xy$ magnetic flux every 50 sweeps.
The dotted line is the leading prediction. Flux states are long-lived and cause a shift of the in-plane plaquette.}
\end{figure}


We can now repeat this construction in the case of $SU(3)$ gauge-symmetry breaking.
Starting from a "cold"\footnote{``Cold'' here stands for a state which minimizes the total lattice action, with non-trivial $P_5$.} configuration, prepare a flux state in some specific $U(1)$ subgroup, in each $xy$ plane:
$\theta_{U(1)} = \frac{2\pi}{L_x L_y}$. Then perform ordinary Monte Carlo sweeps of the $SU(3)$ link variables,
but with an action which is expected to induce gauge-symmetry breaking. In our case, we chose to induce gauge-symmetry
breaking by applying an external potential $h_F {\rm ReTr}P_5 + h_A |{\rm Tr}P_5|^2$, following \cite{Myers:2007vc}.
This is simpler and computationally much cheaper than simulating adjoint fermions.

One effect of the $SU(3)$  Monte Carlo updates is to rotate in $SU(3)$ the $U(1)$ subgroup where the flux was introduced,
independently at each lattice site. To some extent, this local scrambling could be undone by gauge-fixing.
But gauge-fixing is not necessary: what we want to ascertain is whether the $U(1)$ flux is still there.
For that purpose, it is sufficient to monitor the gauge-invariant excess action in the $xy$ planes, namely
$\Delta = \langle {\rm Tr}U_{P_{xz}} \rangle - \langle {\rm Tr}U_{P_{xy}} \rangle$,
just like in the pure $U(1)$ case.

Fig.~4 shows the results of such an experiment in the reconfined phase ($SU(3) \to U(1)\times U(1)$).
On the $y$-axis, $\Delta$ has been normalized to its value in the pure $U(1)$ case. Interestingly,
the excess action depends on the $U(1)$ subgroup where the flux is introduced. The reason is the 
following. In all cases, a $U(1)$ angle $\theta=\frac{2\pi}{L_x L_y}$ is introduced in each $xy$ plaquette.
The corresponding action is $1 - {\rm Tr}(e^{i\theta}) \approx \frac{1}{2}\theta^2$ in the pure $U(1)$ system.
But in the $SU(3)$ system, if the flux is introduced in the $\lambda_3$ subgroup, the corresponding action is
$1 - \frac{1}{3} {\rm Tr}\{{\rm diag}(e^{i\theta}, e^{-i\theta},1)\} \approx \frac{1}{3}\theta^2$.
And if the flux is introduced in the $\lambda_8$ subgroup, the action is
$1 - \frac{1}{3} {\rm Tr}\{{\rm diag}(e^{i\theta}, e^{i\theta},e^{-2i\theta})\} \approx \theta^2$.
Thus, a magnetic flux in the $\lambda_3$ or $\lambda_8$ subgroup incurs an action equal to $2/3$ or
$2$ times that in a $U(1)$ system, respectively. 
This is precisely what Fig.~4 shows, with horizontal dotted lines corresponding to $2/3$ (subgroup $\lambda_3$),
$2$ (subgroup $\lambda_8$) or $14/3$  (2 units of flux in $\lambda_3$ plus 1 unit in $\lambda_8$).
Note that these topological excitations are extremely stable: their sudden
decay after 1000 Monte Carlo sweeps is due to our turning off the external potential which maintained
the reconfined phase. Then, the full $SU(3)$ gauge symmetry is immediately restored, and the $U(1)$ 
fluxes can freely unwind.

\begin{figure}
\centerline{
\includegraphics[width=0.55\textwidth]{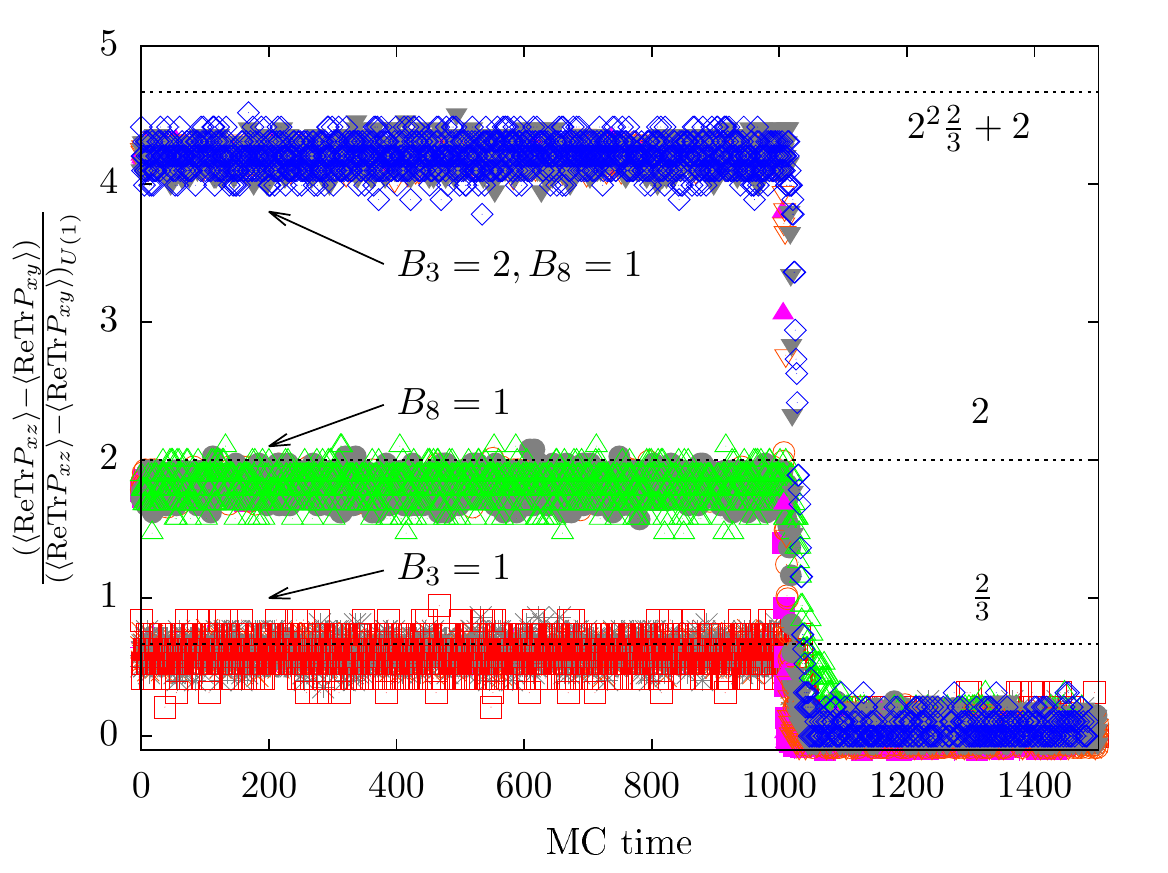}
}
\caption{In an $SU(3)$ system in the reconfined phase ($U(1)\times U(1)$), magnetic flux is introduced in the $xy$ plane. The corresponding
shift in the $xy$ plaquette, normalized to that in a pure $U(1)$ system, is shown for several flux combinations in the $\lambda_3$ and $\lambda_8$
subgroups. The dotted lines show the classical predictions. After 1000 sweeps, the gauge-symmetry breaking potential is turned off: the
full gauge-symmetry is restored and the flux states decay immediately.}
\end{figure}

The next topological defect we have considered is a magnetic monopole, which is visible via its gauge-invariant $3d$ magnetic 
flux through the 6 faces of an elementary cube (see Fig.~5 left). Actually, we are interested in a classical monopole, obtained
by minimizing the action of a $3d$ $U(1)$ lattice of size $L^3$ containing one monopole. To enforce the presence of one
monopole, we introduce a flux $\pi$ through 3 of the faces of the lattice, and choose charge-conjugated periodic boundary
conditions in each of the 3 directions, $U_\mu(x+L) = U_\mu^*(x)$. The resulting construction, Fig.~5 right, is the analogue of the DeGrand-Toussaint
monopole on the left, but on the scale $L$ instead of $a$.  It was already used in \cite{Vettorazzo:2003fg} to measure
the monopole mass, but the numerical results presented there turn out to be incorrect.

Fig.~6 left shows the minimum energy (measured by cooling) of a $U(1)$ magnetic monopole induced by the above boundary conditions,
as a function of the size $L$ of the cubic lattice. To understand its $L$-dependence, consider first a monopole of charge $Q_M =2\pi/e$ in the
continuum. The energy of the magnetic field inside a sphere of radius $R$ is $E(R) = 2\pi \int_0^R dr~r^2~B(r)^2$. It is UV-divergent,
and the lattice spacing $a$ will cutoff the integral and regularize the divergence. In the infrared, since $B(r)=Q_M /(4\pi r^2)$ as
dictated by Gauss' law, one obtains $E(R=\infty) - E(R) = \frac{Q_M^2}{8\pi} \int_R^\infty dr 1/r^2 = \frac{1}{e^2} \frac{\pi}{2R}$.
This calculation is slightly modified in a cubic box with $C$-periodic boundary conditions. The boundary conditions generate an
infinite array of mirror charges, arranged in a cubic array of spacing $L$ and alternating in sign, as in an $Na^+ Cl^-$ crystal.
They interact via a $1/r$ potential, so that 
the energy of the array is proportional to $\alpha_3 = \sum'_{ijk} \frac{(-1)^{i+j+k}}{\sqrt{i^2+j^2+k^2}} = -1.74756..$,
which is called Madelung's constant. The resulting monopole energy correction is 
$E(R=\infty) - E(R) = \frac{1}{e^2} \frac{\alpha_3}{2} \frac{\pi}{L} = \frac{2.745..}{L}$. This is precisely the $1/L$ dependence
seen in Fig.~6 left. Since its origin is infrared, this term is universal, i.e. independent of the lattice action considered.
The leading term of course depends on the form of the ultraviolet cutoff, and thus is action-dependent.
The additional, $1/L^3$, tiny corrections come from $(a/L)^2$ lattice corrections to the continuum Coulomb potential.

Now, as in the case of $U(1)$ fluxes, we can introduce a $U(1)$ monopole in a subgroup of an $SU(3)$ configuration.
If the gauge symmetry is broken by the external Polyakov loop potential, the $U(1)$ monopole is stable, and we can
measure its energy by cooling. The results are shown Fig.~6 right. First, the monopole energy depends on the $U(1)$
subgroup chosen, just like for fluxes. This fact was first noticed in Ref.~\cite{Cea:2000zr} which used Maximal Abelian gauge and 
Abelian projection to isolate the monopoles. Here, we obtain precise values for the monopole energies in the thermodynamic
limit: contrary to the energies of flux states, the monopole energies are not obtained by applying a simple factor to
the $U(1)$ case, because the UV-regularization of the monopole field differs in the different subgroups.
Nevertheless, the coefficient of the $1/L$ correction, which comes from IR effects, varies as for flux states:
it is $2/3$ and $2$ times the $U(1)$ value for subgroups spanned by $\lambda_3$ and $\lambda_8$, respectively.
Moreover, the $1/L^3$ coefficients scale in the same way, since they are all caused by the same lattice distortion
of the Coulomb potential.


The topological excitations considered here, Abelian fluxes and monopoles, are appropriate for diagnosing
gauge-symmetry breaking to a $U(1)$ subgroup. To diagnose gauge-symmetry breaking to an $SU(2)$ subgroup,
one should monitor the stability of a 't Hooft-Polyakov monopole. \linebreak This next step is under investigation.
Finally, it is clear that our approach can be used without change to diagnose gauge-symmetry breaking
in an ordinary gauge-Higgs system. Note that our construction is completely non-local, so that it does not
contradict the Fradkin-Shenker argument against the existence of an order parameter distinguishing 
the Higgs and the confining regimes.

\begin{figure}
\vspace*{-4mm}
\centerline{
\hspace*{10mm}
\includegraphics[width=0.20\textwidth]{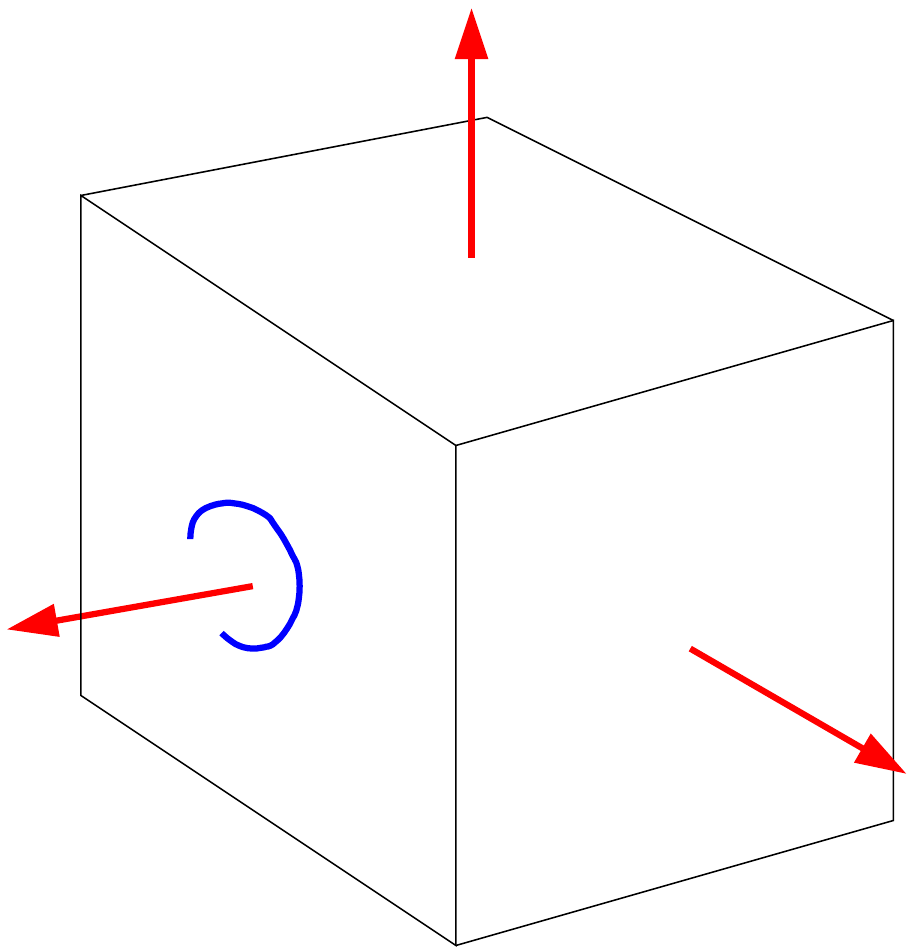}
\put(-26,39){\large $a$}
\hspace*{5mm}
\begin{minipage}{0.25\textwidth}
\vspace*{-25mm}
$\sum_{6~{\rm plaq}} \theta_{\rm plaq} = 2\pi$
\end{minipage}
\includegraphics[width=0.20\textwidth]{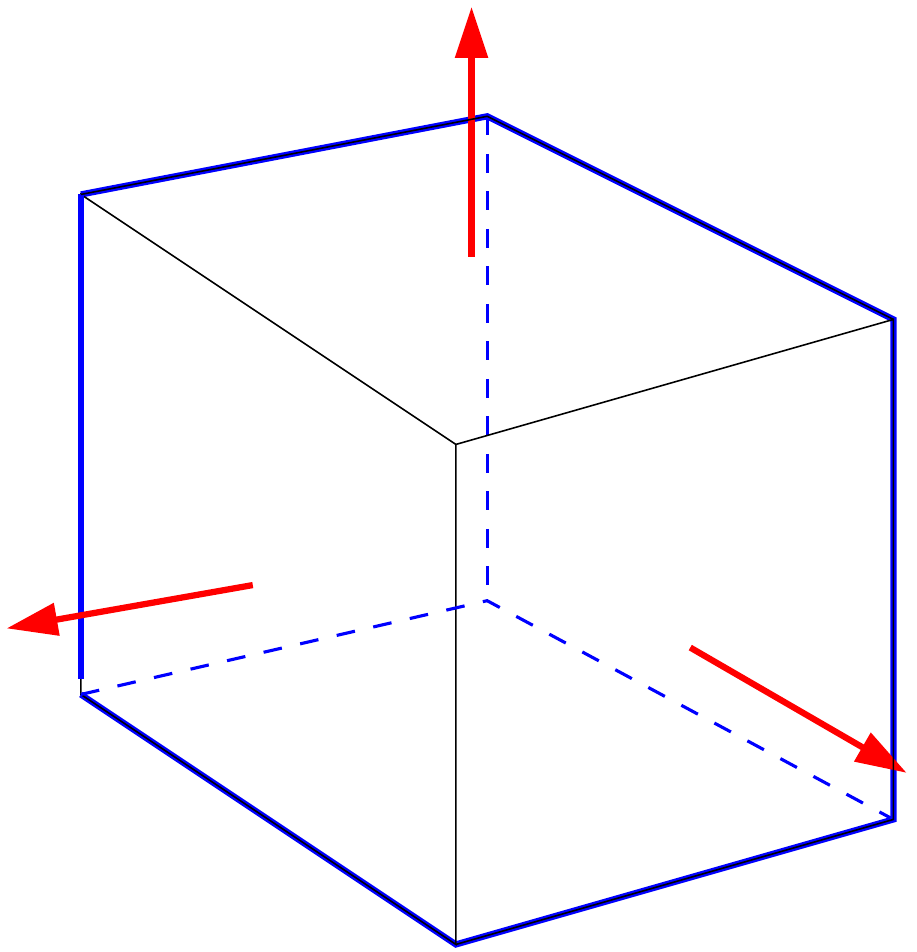}
\put(-24,37){\large $L$}
\hspace*{5mm}
\begin{minipage}{0.25\textwidth}
\vspace*{-25mm}
$\sum_{3~{\rm planes}} \theta_{\rm plane} = \pi$
\end{minipage}
}
\caption{({\em left}) DeGrand-Toussaint magnetic monopole in an elementary cube of size $a$. ({\em right}) The same construction on the scale $L$
of the whole lattice ensures the presence of a magnetic monopole somewhere inside. Charge-conjugated boundary conditions are required to 
obtain non-zero fluxes at the boundary.}
\vspace*{-3mm}
\end{figure}

\begin{figure}
\centerline{
\includegraphics[width=0.5\textwidth]{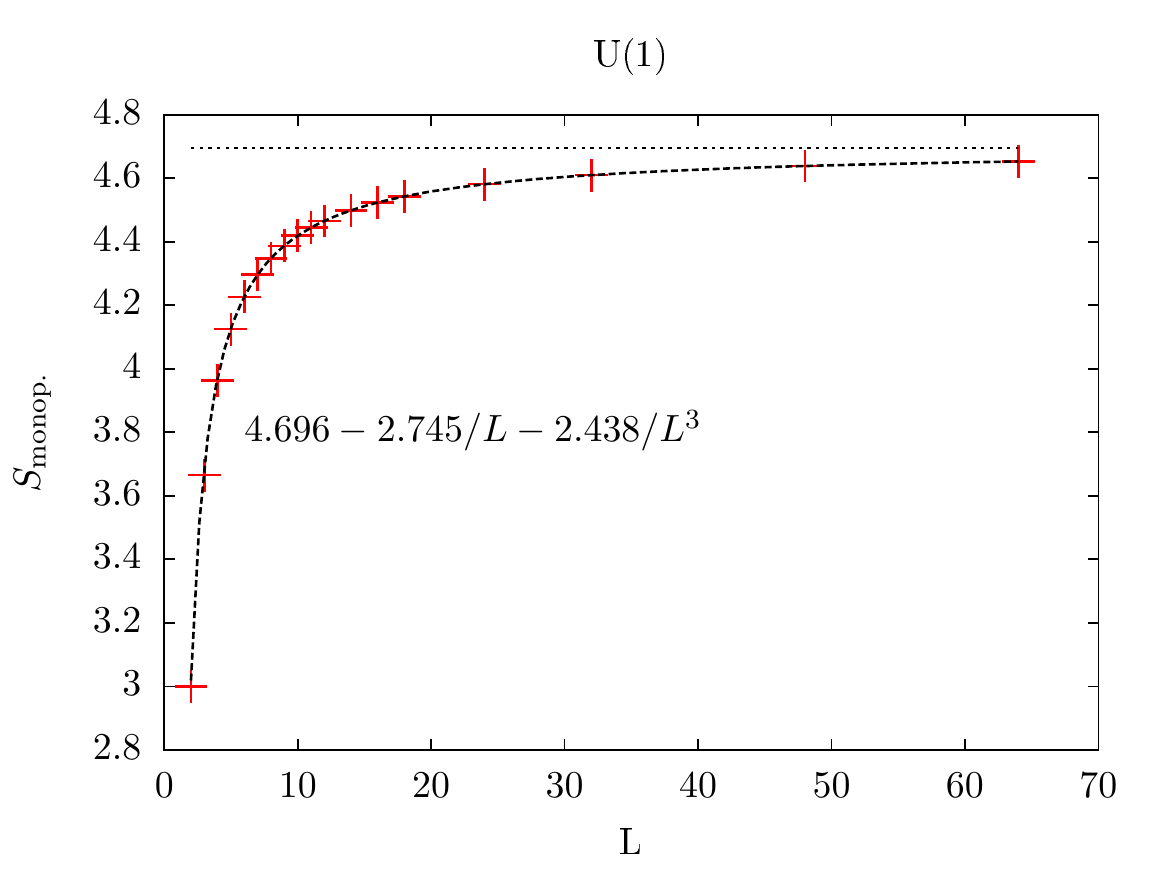}
\hspace*{2mm}
\includegraphics[width=0.5\textwidth]{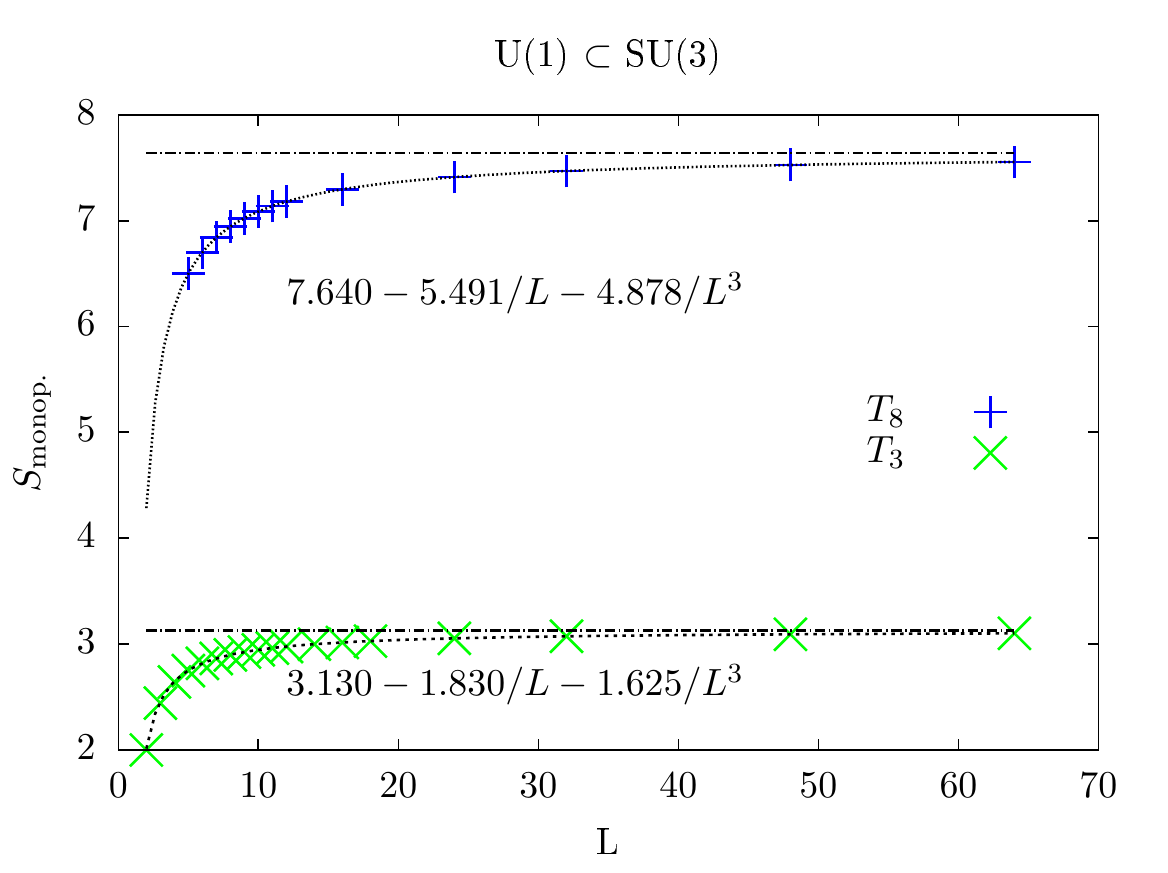}
}
\vspace*{-3mm}
\caption{({\em left}) Minimum action of a $U(1)$ magnetic monopole as a function of the size $L$ of the cubic lattice. The $1/L$ correction is caused
by the cubic array of image-charges with alternating signs, and its magnitude is exactly given by Madelung's constant. ({\em right}) Same, for 
a $U(1)$ magnetic monopole in the $\lambda_3$ or $\lambda_8$ sector of an $SU(3)$ system in the reconfined ($U(1)\times U(1)$) phase.
The two types of monopoles have different masses. The $1/L$ correction is given by the correspondingly rescaled Madelung constant. }
\vspace*{-3mm}
\end{figure}



\end{document}